\documentclass{epl2}
\usepackage{epsfig}
\usepackage{graphicx}
\usepackage{amsmath}
\usepackage{amssymb}

\begin{document}

\title{The influence of temporal coherence on the dynamical Casimir effect}

\author{J.T. Mendon\c{c}a\inst{1,\dag} \and G. Brodin\inst{2} \and M. Marklund\inst{2}}
\shortauthor{J.T. Mendon\c{c}a \etal}

\institute{
\inst{1} IPFN, Instituto Superior T\'{e}cnico,1049-001 Lisboa, Portugal \\
\inst{2} Department of Physics, Ume{\aa} University, SE--901 87 Ume{\aa}, Sweden \\
\inst{\dag} \emph{E-mail:} titomend@ist.utl.pt}

\abstract{
We study the dynamical Casimir effect in the presence of a finite coherence time, which is associated with a finite quality factor of the optical cavity. We use the time refraction model, where a fixed cavity with a modulated optical medium, replaces the empty cavity with a vibrating mirror. Temporal coherence is described with the help  of cavity quasi-mode operators. Asymptotic expressions for the number of photon pairs generated from vacuum are derived.
}

\maketitle

\section{Introduction}

The dynamical Casimir effects is an instability of the electrodynamical vacuum, which has received considerable attention in recent years (see for a recent review\cite{dodrev}). For an empty optical cavity with a vibrating mirror, with no losses, it leads to the exponential growth of trapped photons, if the oscillating frequency of the mirror is equal to twice the photon frequency. Its main interest is related with the fact that it can occur for arbitrarily low photon energies, thus stimulating the hope of observation of quantum vacuum properties in small scale experiments. 

Another approach to this problem is offered by the concept of time refraction \cite{mendpra}, which is associated to the temporal change of the optical properties of a dielectric medium. The similarities between time refraction and the dynamical Casimir effect were recently explored by us \cite{mendepl}, where the empty cavity with a vibrating mirror was replaced by a static cavity with a time dependent optical medium. The equivalence between the two configurations was demonstrated, and again, in the absence of losses, an exponential growth of the photon number inside the cavity was obtained. We have also discussed the influence of losses, which are always present in any optical cavity. These losses could be due to dissipation in the optical medium, or to the imperfections of the cavity mirrors. The existence of losses can therefore be associated with a finite quality factor $Q$, which on the other hand, characterizes the finite lifetime of the trapped photons. This will lead to a finite coherence time for the quantum field states, as shown here.

Our phenomenological model for dissipation \cite{mendepl} had predicted that the exponential growth of the photon number would only take place for a photon emission rate larger than the photon losses. This agrees qualitatively with other models considering the dynamical Casimir effects in the presence of losses \cite{kim,dod98,schaller,dodrev}. In such models, no asymptotic value for the emitted number of photons would be attained for very long times. 

In contrast, recent work \cite{lamb,lamb2}, has shown that an asymptotic equilibrium value could be derived, for low excitation rates. Such a value would result from a balance between the emission process and the photon losses. However, with this alternative theoretical model, no time dependent result can be derived, giving us no information on the dynamics of the photon emission process. Even if the results were in a fair agreement with ours for a single cavity mode, the question remains that an asymptotic state was shown to exist in contrast with our previous derivation.

Here we return to the problem of the cavity losses, and show that, by using our time refraction approach it is possible to derive a temporal photon emission rate which tends, for very long times to a constant asymptotic value. Under appropriate assumptions, it is possible to show that this asymptotic value coincides with the one derived by \cite{lamb}, thus establishing a bridge between the two theoretical models and showing that the two approaches can indeed be reconciled into a single unified picture. The key element in this unified picture is provided by the existence of  a finite coherence time of the photon quantum states associated to the cavity losses.

In this work, we extend our time refraction model to consider the presence of losses, by using the concept of cavity quasi-mode, which was first developed in the frame of laser cavities \cite{fox}, and then extended and applied by many authors \cite{barnett,gea,law}. This concept, and the corresponding field operators, are described in Section 2. Time refraction of a cavity quasi-mode is then discussed in Section 3, where an expression for the time dependent photon number creation is established and its main physical consequences are discussed. Finally, in Section 4, we state our conclusions. 

\section{Cavity quasi-modes}

We start by considering a static dielectric medium in a one-dimensional optical cavity, which is defined by two parallel mirrors, placed at positions $z=0$ and $z=L$. The transverse mode structure is somewhat marginal to our present discussion and is ignored here. It can however be easily accommodated in the formalism, and it will be briefly discussed at the conclusions. The classical electric field describing the modes in the cavity can be written as
\begin{equation}
E(z, t) = \sum_m E_m \sin (k_m z) \exp (- i \omega_m t) + c.c.
\label{eq:2.1} \end{equation}
with the mode wavenumbers and frequencies
\begin{equation}
k_m = 2 \pi \frac{m}{L} \; , \quad \omega_m = \frac{k_m c}{\sqrt{\epsilon (\omega_m)}}
\label{eq:2.1b} \end{equation}
where $m$ is an integer, $c$ the velocity of light in vacuum, and $\epsilon (\omega)$ is the dielectric function of the medium, assumed generically to depend on the field frequency $\omega$. This is replaced, in a quantum description, by the electric field operator
\begin{equation}
E (z, t) = \sum_m \left[ E_m (t) \exp (i k_m z) + h.c. \right]
\label{eq:2.2} \end{equation}
with 
\begin{equation}
E_m (t) = i C_m \left[ a_m (t)- a^\dag_{-m} (t) \right]
\label{eq:2.2b} \end{equation}
where $C_m = \sqrt{\hbar \omega_m / 2 \epsilon (\omega_m)}$ is an appropriate normalizing factor. It should be noticed that the standing wave field described by equation (\ref{eq:2.1}) results from the superposition of two counter-propagating photon states, as explicitly represented in equations (\ref{eq:2.2})-(\ref{eq:2.2b}). For each cavity mode $m$, the destruction and creation operators can be defined as
\begin{equation}
a_m (t) = A_m \exp (-i \omega_m t) \; , \quad a^\dag_{-m} = A^\dag_{-m} \exp (i \omega_m t)
\label{eq:2.3} \end{equation}
Let us now assume that these modes correspond to a lossy cavity, or in other words, a cavity with a finite quality factor $Q_m$. This is equivalent to introduce a finite spectral width for the cavity mode, $\gamma_m = \omega_m / Q_m$. In this case, we can still use a similar mode description, but where the cavity modes are not really electric field modes, and can be considered as cavity quasi-modes. The operators $a_m (t)$ and $a^\dag_{-m} (t)$ now become {\sl effective mode operators}, also called {\sl quasi-mode operators}, resulting from the superposition of an infinite number of field modes with nearly equal frequencies $\omega \sim \omega_m$ \cite{law}. These effective operators, can then be stated as
\begin{equation}
a_m (t) = \int g_m (\xi) \tilde{a}_m (\xi, t) \frac{d \xi}{2 \pi} \; , \quad
a^\dag_{-m} (t) = \int g^*_m (\xi) \tilde{a}^\dag_{-m} (\xi, t) \frac{d \xi}{2 \pi}
\label{eq:2.4} \end{equation}
where $\xi = (\omega - \omega_m)$ is an internal frequency variable, and $g_m (\xi)$ is the spectral shape function to be defined later. It is obvious from here that the effective mode operators $a_m (t)$ and $a^\dag_m (t)$ can be considered as external or macroscopic operators, whereas the field modes $\tilde{a}_m (\xi, t)$ and $\tilde{a}^\dag_{-m} (\xi, t)$ are the internal or microscopic ones. These internal operators can similarly be decomposed in amplitude and phase as
\begin{equation}
\tilde{a}_m (t) = \tilde{A}_m (\xi) \exp [-i \phi (t)] \; , \quad \tilde{a}^\dag_{-m} = \tilde{A}^\dag_{-m} \exp [i \phi (t)]
\label{eq:2.5} \end{equation}
where the phase varies in time according to $\phi (t) = \omega t \equiv (\omega_m + \xi) t$. Comparing this with equation (\ref{eq:2.3}), we conclude that
\begin{equation}
A_m = \int g_m (\xi) \tilde{A}_m (\xi) \exp (-i \xi t) \frac{d \xi}{2 \pi} \; , \quad 
A^\dag_{-m} = \int g^*_m (\xi) \tilde{A}^\dag_{-m} (\xi) \exp (i \xi t) \frac{d \xi}{2 \pi} 
\label{eq:2.6} \end{equation}
For each internal field mode $\xi$ we can define a photon number operator, in the usual way, as
\begin{equation}
\tilde{N} (\xi) = \tilde{a}^\dag_m (\xi, t) \tilde{a}_m (\xi, t) = \tilde{A}^\dag_m (\xi) \tilde{A}_m (\xi)
\label{eq:2.7} \end{equation}
Using equations (\ref{eq:2.5}), we can then define a photon number operator for the effective cavity mode $m$, as
\begin{equation}
N_m (t) = \int \frac{d \xi}{2 \pi} \int \frac{d \xi'}{2 \pi} g^*_m (\xi) g_m (\xi') \tilde{A}^\dag_m (\xi) \tilde{A}_m (\xi') 
\label{eq:2.8} \end{equation}
At this point it should be noticed that the spectral shape function $g_m (\xi)$ can be chosen in such a way as to satisfy the normalization condition
\begin{equation}
\int |g_m (\xi)|^2 \frac{d \xi}{2 \pi} = 1
\label{eq:2.9} \end{equation}
This particular choice will enable the effective mode operators to obey the commutation relation $[a_m, a^\dag_{m'} ]Ê=\delta_{m, m'}$, given the usual commutation relations valid for the internal field mode operators
\begin{equation}
[\tilde{a}_m (\xi), \tilde{a}^\dag_{m'} (\xi') ] = \delta_{m, m'} \delta (\xi - \xi')
\label{eq:2.10} \end{equation}
as demonstrated by previous work on lossy cavities \cite{law}. Until now, we have assumed the shape function $g_m (\xi)$ as an arbitrary complex function, obeying the normalization condition (\ref{eq:2.8}). But a natural choice, which describes a Lorentzian mode profile with half-width $\gamma_m$ is given by
\begin{equation}
g_m (\xi) = \frac{\sqrt{2 \gamma_m}}{(\xi^2 + \gamma_m^2)^{1/2}}
\label{eq:2.11} \end{equation}
It is obvious that the particular case $|g_m (\xi)|^2 = 2 \pi \delta (\xi)$ would lead to $A_m = \tilde{A}_m$ and $A^\dag_{-m} = \tilde{A}^\dag_{-m}$, and corresponds to the ideal case of a perfect cavity, valid in the limit $\gamma_m \rightarrow 0$.

Before concluding this introductory section on the cavity quasi-modes let us consider the field Hamiltonian $H$. Following the usual definition, we can obviously write $H = \sum_m H_m$, with
\begin{equation}
H_m = \hbar \int (\omega_m + \xi) \left[ \tilde{N}_m (\xi) + \frac{1}{2} \right] d \xi
\label{eq:2.12} \end{equation}
This allows us to derive the Heisenberg equations for the field operators, as
\begin{equation}
\frac{d}{d t} \tilde{a}_m (\xi, t) = \frac{1}{i \hbar} \left[ \tilde{a}_m (\xi, t), H_m \right] = - i (\omega_m + \xi) \tilde{a}_m (\xi, t)
\label{eq:2.12b} \end{equation}
with a similar expression for $\tilde{a}^\dag_{-m} (\xi, t)$. This could also be recovered directly from equations (\ref{eq:2.5}).

\section{Time refraction of a quasi-mode}

Let us now consider the temporal variation of the optical properties of the medium inside the cavity. This can be described by assuming a varying dielectric function $\epsilon (\omega_m, t)$. Given the fact that $k_m$ is fixed, as imposed by the fixed boundaries, the frequency $\omega_m$ has to change in time, in order to satisfy the dispersion relation (\ref{eq:2.1b}) for all times. This situation corresponds to a time refraction inside the cavity, as previously discussed by us \cite{mendepl,mendpra}. The cavity quasi-modes can now be defined as
\begin{equation}
a_m (t) = A_m (t) \exp [-i \phi_m (t)] \; , \quad a^\dag_{-m} = A^\dag_{-m} (t) \exp [i \phi_m (t)]
\label{eq:3.1} \end{equation}
where, in contrast with equation (\ref{eq:2.3}), the amplitudes are now time dependent operators, and the phase is determined by
\begin{equation}
\phi_m (t) = \int^t \omega_m (t') d t«= k_m c \int^t \frac{d t'}{n (t')}
\label{eq:3.1b} \end{equation}
where $n (t) = \sqrt{ \epsilon (\omega_m, t)}$ is the refractive index of the non-stationary medium. As for the internal field modes, they can still be described by equations (\ref{eq:2.5}), but with time varying amplitudes $\tilde{A}_m (\xi, t)$ and $\tilde{A}^\dag_{-m} (\xi, t)$, and with a new phase defined by $\phi_\xi (t) = \phi_m (t) + \xi t$. 

We can now use the well known properties of time refraction \cite{mendpra,mendbook}. As previously shown, in a time varying medium, the field mode operators are coupled by the evolution equations
\begin{equation}
\frac{d}{d t} \tilde{a}_m = - i \omega \tilde{a}_m + f (t) \tilde{a}^\dag_{-m} \; , \quad
\frac{d}{d t} \tilde{a}^\dag_{-m} =  i \omega \tilde{a}^\dag_{-m} + f (t) \tilde{a}_m
\label{eq:3.2} \end{equation}
where $\omega = \omega_m + \xi$, and $f (t)$ is a real function defining the temporal evolution of the refractive index 
\begin{equation}
f (t) = \frac{1}{2 n} \frac{d n}{d t}
\label{eq:3.2b} \end{equation}
We can also rewrite equations (\ref{eq:3.2}) in terms of the field amplitude operators, as
\begin{equation}
\frac{d}{d t} \tilde{A}_m = \nu_\xi (t) \tilde{A}^\dag_{-m} \; , \quad
\frac{d}{d t} \tilde{A}^\dag_{-m} =  \nu^*_\xi (t) \tilde{A}_m
\label{eq:3.3} \end{equation}
where for simplicity, the dependence of the operators on $\xi$ was not explicitly stated, and where we have defined 
\begin{equation}
\nu_\xi (t) = f (t) \exp [ 2 i \phi_\xi (t) ]
\label{eq:3.3b} \end{equation}
Integration of equations (\ref{eq:3.3}) leads to the following solution
\begin{eqnarray}
\tilde{A}_m (t) &=& \cosh [r (\xi, t)] \tilde{A}_m (0) + \sinh [r (\xi, t)] \tilde{A}^\dag_{-m} (0) e^{i \delta_\xi}
\nonumber \\
\tilde{A}^\dag_{-m} (t) &=& \cosh [r (\xi, t)] \tilde{A}^\dag_{-m} (0) + \sinh [r (\xi, t)] \tilde{A}_m (0) e^{-i \delta_\xi} 
\label{eq:3.4} \end{eqnarray}
where we have introduced the squeezing function
\begin{equation}
r (\xi, t) = \int^t | \nu_\xi (t') | dt' = \int^t f (t') e^{2 i \xi t'} e^{2i \phi_m (t') - i \delta_\xi} d t'
\label{eq:3.4b} \end{equation}
The phase $\delta_\xi$ is defined by $\nu_\xi (t) = | \nu_\xi (t) | \exp (i \delta_\xi)$, and will not contribute to the mean photon number, to be considered below. We can therefore take it as $\delta_\xi = 0$. An alternative and equivalent option, would be to incorporate this phase in the definition of the operator $\tilde{A}^\dag_{-m}$.
We can now calculate, for each mode $\omega$ or $\xi$, the mean photon number, as $< \tilde{N}_m (\xi, t) > = < 0 | \tilde{a}^\dag_m \; \tilde{a}_m | 0 >$, where $|0>$ is the unperturbed vacuum state. This leads to
\begin{equation}
< \tilde{N}_m (\xi, t) > = \sinh^2 [ r (\xi, t)]
\label{eq:3.5} \end{equation}
It should be noticed that the same number of photons will be emitted in the opposite direction, $ < \tilde{N}_{-m} (\xi, t) > = < \tilde{N}_m (\xi, t) >$, in order to guarantee that the total momentum of the electromagnetic vacuum is conserved and remains equal to zero. This expression therefore states the number of photon pair emitted from vacuum.
According to equation (\ref{eq:2.8}), this then leads to the following result for the mean photon number associated with the cavity quasi-mode $m$
 \begin{equation}
< N_m (t) > = 2 \gamma_m \int \frac{\sinh^2 [ r (\xi, t)]}{(\xi^2 + \gamma_m^2)} \; \frac{d \xi}{2 \pi}
\label{eq:3.6} \end{equation}
where a Loremtzian profile was assumed. This is our man result, which has important physical consequences to be discussed next. In particular, it will lead to the occurrence of a saturation of the dynamical Casimir instability, due to the existence of a finite coherence time for the cavity fields, charaterized by $1 / \gamma_m$.

Let us first assume a sinusoidal perturbation of the refractive index of the medium, such that 
\begin{equation}
n (t) = n_0 + \delta n (t) \; , \quad \delta n (t) = n_0 \epsilon \sin (\Omega t)
\label{eq:3.7} \end{equation}
where $\Omega$ is the driving frequency, and the amplitude perturbation is assumed very small, $\epsilon \ll 1$. As discussed previously by us \cite{mendepl}, this is optically equivalent to an empty cavity with an oscillating length, $L (t) = L_0 + \delta L (t)$, where $\delta L (t) = L_0 \delta n (t) / n_0$. In this case, we can write
\begin{equation}
f (t) \simeq \frac{\epsilon \Omega}{2} \cos (\Omega t)
\label{eq:3.8} \end{equation} 
\begin{equation}
\phi_m (t) = \int^t \omega_m (t') d t' \simeq \omega_0 t + \epsilon \frac{\omega_0}{\Omega} \cos (\Omega t)
\label{eq:3.8b} \end{equation}
where we have used $\omega_0 = k_m c / n_0$. This leads to a phase factor of the form
\begin{equation}
e^{2i\phi_m (t)} = \sum_{l=-\infty}^\infty i^l J_l (\alpha) e^{i(2 \omega_0 + l \Omega) t}
\label{eq:3.8c} \end{equation}
where $\alpha = 2 \epsilon \omega_0 / \Omega$, and $J_l$ are Bessel functions of order $l$. Replacing this in the expression for the squeezing function, we get
\begin{equation}
r (\xi, t) = \frac{\epsilon \Omega}{4} \sum_l i^l J_l (\alpha) \int_0^t \left[ e^{i \omega_+ t'} + e^{i \omega_- t'} \right] e^{2 i \xi t'} d t'
\label{eq:3.9} \end{equation}
where 
\begin{equation}
\omega_\pm = 2 \omega_0 + (l \pm 1) \Omega
\label{eq:3.9b} \end{equation}
In order to maximize the value of $r (\xi, t)$, we have to choose the driving frequency $\Omega$ such that one of the frequencies $\omega_\pm$ is made equal to zero. This corresponds to
\begin{equation}
\Omega = \Omega_l \equiv - \frac{2 \omega_0}{l \pm 1}
\label{eq:3.10} \end{equation}
which implies that
\begin{equation}
\alpha = \alpha_l \equiv - \epsilon ( l \pm 1)
\label{eq:3.10b} \end{equation}
We generally expect to have $\alpha \ll 1$, even for large values of $|l| \gg 1$. This allows us to use the approximate expression for the Bessel functions, $J_l (\alpha) 
\simeq \alpha^l / 2^l l!$. It becomes apparent that the dominant term in equation (\ref{eq:3.9} will correspond to $l = 0$, although this is not convenient experimentally because it implies a large driving frequency $\Omega = 2 \omega_0$. The squeezing function then becomes
\begin{equation}
r (\xi, t) = \nu_0 \int_0^t \exp (2 i \xi t' ) d t'
\label{eq:3.11} \end{equation}
with $\nu_0 = \epsilon \omega_0 / 2$. Let us first consider the case $|r (\xi, t)| \ll 1$. This allows us to simplify equation (\ref{eq:3.6}), which can now be apprimately written as
\begin{equation}
< N_m (t) > =  \nu_0^2 \int \frac{d \xi}{2 \pi} \frac{2 \gamma_m}{(\xi^2 + \gamma_m^2)} \left[ \int_0^t e^{2i\xi t'} d t' \right]^2
\label{eq:3.12} \end{equation}
This can also be stated as
\begin{equation}
< N_m (t) > =  \nu_0^2 \int_0^t d t' \int_0^t d t'' \int \frac{d \xi}{2 \pi} \frac{2 \gamma_m}{(\xi^2 + \gamma_m^2)} e^{2i\xi (t' + t'')}
\label{eq:3.12b} \end{equation}
Noting that
\begin{equation}
\int_{-\infty}^\infty \frac{e^{-i \omega t}}{(\omega^2 + \gamma^2)} \frac{d \omega}{2 \pi} = \frac{1}{2 \gamma} \exp (- \gamma | t|)
\label{eq:3.13} \end{equation}
integration over $\xi$ can easily be performed, leading to
\begin{equation}
< N_m (t) > =  \nu_0^2 \int_0^t d t' \int_0^t d t'' \int e^{- \gamma_m |t' + t''|} = \nu_0^2 \left[ \int_0^t e^{- \gamma_m t'} d t' \right]^2
\label{eq:3.14} \end{equation}
We can finally write
\begin{equation}
< N_m (t) > =   \frac{\nu_0^2}{\gamma_m^2} \left( 1 - e^{- \gamma_m t} \right)^2
\label{eq:3.15} \end{equation}
We should notice here that, for $t \gg \tau$, where $\tau = 1 / \gamma_m$ can be define as the field coherence time for the cavity mode $m$, the number of photon pairs determined by this expression attains the asymptotic value of
\begin{equation}
< N_m (\infty) > =   \nu_0^2 \; \tau^2
\label{eq:3.15b} \end{equation}
This eactly coincides with the steady state result derived from a completely different method by Lambrecht et al \cite{lamb}, for a single cavity mode, thus showing that our time refraction approach to the dynamical Casimir effect is indeed capable to describe the temporal evolution of the number of photon pairs created from vacuum, and to establish an asymptotic state for a lossy cavity. The possible existence of different modes contributing to the effect will be discussed below. On the other end, equation (\ref{eq:3.15}) also shows that, in the oposite limit of short times, $t \ll \tau$, we get $< N_m (t) > =   \nu_0^2 \; t^2$, which agrees with our previous result in the same limit \cite{mendepl}. We can therefore state that the existence of an asymptotic steady state is provided by a finite decoherence time $\tau$ for the field in the cavity.

\begin{figure}
\onefigure[width=.8\columnwidth]{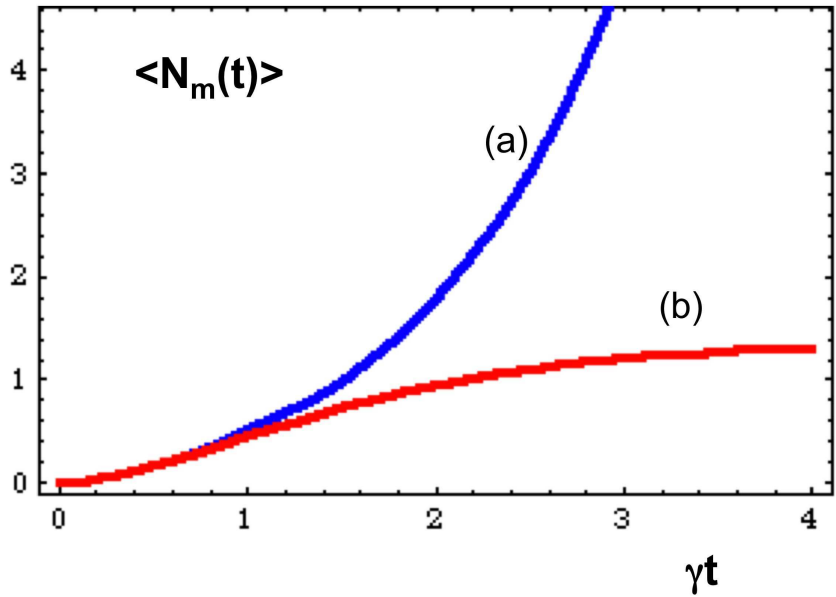}
\caption{{\sl Number of photon pairs created from vacuum: (a) as determined by the phenomenological model $< N_m (t) > = \sinh^2 (\nu_0 t) \exp (- \gamma_m t)$, where exponential growth always occurs for $\nu_0  > \gamma_m$; (b) as determined by equation (\ref{eq:3.16}), which shows the existence of a asymptotic saturation  level.}}
\label{fig}
\end{figure}

The above discussion is only valid for $\nu_0 \tau \ll 1$, compatible with the assumption of $| r (\xi, t) | \ll 1$. which led to equation (\ref{eq:3.15}). But, in order to observe photon creation, and to get $< N_m (\infty) > \; \geq 1$, we should be able to consider a more general case. Let us then determine the number of photons $< N_m (t) >$ for an arbitrary squeezing factor.

In the general case, where $| r (\xi, t)|$ can take values much larger than one, we can develop the $sinh^2$ function in a power series, and for each term of this series, perform the integration over $\xi$ in the way described above. After taking the sum of the resulting series, the final result is
\begin{equation}
< N_m (t) > =   \sinh^2 \left[ \frac{\nu_0}{\gamma_m} \left( 1 - e^{- \gamma_m t} \right)\right]
\label{eq:3.16} \end{equation}
We can see here that the initial exponential growth associated with the dynamical instability is limited by a coherence time. This result is therefore qualitatively different from previous estimates based on phenomenological models for the cavity losses, which always predicted exponential growth, as long as $\nu_0 > \gamma_m$.  This is illustrated in the figure. Here again the same condition is required to excite a large number of photons, but this number tends to saturate for long times,  $t \gg \tau$. The asymptotic value is now given by
\begin{equation}
< N_m (\infty) > =   \sinh^2 \left( \frac{\nu_0}{\gamma_m} \right)
\label{eq:3.17} \end{equation}
which generalizes the asympotic expression derived by \cite{lamb}.

\section{Conclusions}

In this work we have considered the emission of photon pairs due to the occurrence of time refraction in a lossy cavity with an oscillating dielectric medium. This configuration is equivalent to that of a dynamical Casimir effect in a lossy cavity with an oscillating length. This extends our previous work \cite{mendepl} to the case where the coherence time of the electromagnetic field in the cavity is finite, as imposed by a finite lifetime of the confined photons. Such an extension was based on the use of cavity quasi-mode or effective mode operators.

We have shown that our time refraction approach to the dynamical Casimir effect is able to describe the temporal evolution of the photon number emission from vacuum. In particular, it was shown that the existence of a finite coherence time leads to a photon number saturation. Our result for this saturation value is valid for arbitrarily large squeezing parameters, and reduces to the value previously found from a completely different method \cite{lamb} for a single cavity mode, for  the particular case of weak squeezing. However, in contrast with \cite{lamb}, we only consider processes associated with photon pair production with equal frequencies $\omega = \Omega/2$, and we exclude processes associated with unbalanced photon frequencies $\omega+\omega' = \Omega$. The reason is that these unbalanced emission processes violate the conservation of total vacuum momentum. Of course, we can always excite more than one cavity mode. First of all, we have modes with equal axial wavenumber, $k_m$, but different transverse structure. In general terms, this transverse optical structure can be decomposed into a sum of orthogonal Guass-Hermite or Guass-Laguerre states \cite{lasers}. These modes have a slightly different dispersion relation, and a bunch of nearest cavity modes can eventually be excited inside a short frequency band, as imposed by the oscillation of the optical medium. But the losses for the higher transverse modes tend also to be larger. On the other hand, different axial modes $k_m$ can also be excited, as defined by the condition (\ref{eq:3.10}). But, under plausible experimental conditions, only one cavity mode will dominate, as considered here.

\end{document}